\documentclass[aps,prl,twocolumn,superscriptaddress,showpacs]{revtex4}
\usepackage{amsfonts}
\usepackage{amssymb}
\usepackage{amsmath}
\usepackage{graphicx}% Include figure files
\usepackage{dcolumn}% Align table columns on decimal point
\usepackage{bm}% bold math
\usepackage{flafter}
\usepackage{epstopdf}

\begin{document}

\title{Orbital selective Fermi surface shifts and mechanism of high T$_c$ superconductivity in correlated {\it A}FeAs ({\it A}=Li,Na)}

\author{Geunsik Lee }
\email{maxgeun@postech.ac.kr}
\affiliation{Department of Chemistry, Pohang University of Science and Technology, Pohang 790-784, Republic of Korea}
\affiliation{Center for Superfunctional Materials, Department of Chemistry, Pohang University of Science and Technology, Pohang 790-784, Republic of Korea}
\author{Hyo Seok Ji}
\affiliation{Department of Chemistry, Pohang University of Science and Technology, Pohang 790-784, Republic of Korea}
\author{Yeongkwan Kim}
\affiliation{Institute of Physics and Applied Physics, Yonsei University, Seoul, Republic of Korea }
\author{Changyoung Kim}
\affiliation{Institute of Physics and Applied Physics, Yonsei University, Seoul, Republic of Korea }
\author{Kristjan Haule}
\affiliation{Department of Physics and Astronomy, Rutgers University, Piscataway, New Jersey 08854, United States}
\author{Gabriel Kotliar}
\affiliation{Department of Physics and Astronomy, Rutgers University, Piscataway, New Jersey 08854, United States}
\author{Bumsung Lee}
\affiliation{Department of Physics and Astronomy, Seoul National University, Seoul 151-747, Republic of Korea}
\author{Seunghyun Khim}
\affiliation{Department of Physics and Astronomy, Seoul National University, Seoul 151-747, Republic of Korea}
\author{Kee Hoon Kim}
\affiliation{Department of Physics and Astronomy, Seoul National University, Seoul 151-747, Republic of Korea}
\author{Kwang S. Kim}
\affiliation{Center for Superfunctional Materials, Department of Chemistry, Pohang University of Science and Technology, Pohang 790-784, Republic of Korea}
\affiliation{Department of Physics, Pohang University of Science and Technology, Pohang 790-784, Republic of Korea}
\author{Ki-Seok Kim}
\affiliation{Asia Pacific Center for Theoretical Physics, Pohang University of Science and Technology, Pohang 790-784, Republic of Korea}
\affiliation{Department of Physics, Pohang University of Science and Technology, Pohang 790-784, Republic of Korea}
\author{Ji Hoon Shim}
\email{jhshim@postech.ac.kr}
\affiliation{Department of Chemistry, Pohang University of Science and Technology, Pohang 790-784, Republic of Korea}
\affiliation{Department of Physics, Pohang University of Science and Technology, Pohang 790-784, Republic of Korea}

\date{\today}

\begin{abstract}
%Based on fully self-consistent density functional theory (DFT) and
%dynamical mean field theory (DMFT) methods,
Based on the dynamical mean field theory (DMFT) and angle resolved photoemission spectroscopy (ARPES),
we have investigated the mechanism of high $T_c$ superconductivity in stoichiometric LiFeAs.
The calculated spectrum is in excellent agreement with the observed ARPES measurement.
The Fermi surface (FS) nesting, which is predicted in the conventional density functional theory method, is suppressed due to the orbital-dependent correlation effect with the DMFT method.
We have shown that such marginal breakdown of the FS nesting
is an essential condition to the spin-fluctuation mediated superconductivity,
while the good FS nesting in NaFeAs induces a spin density wave ground state.
Our results indicate that fully charge self-consistent description of the correlation effect
is crucial in the description of the FS nesting-driven instabilities.

\end{abstract}


\maketitle
%(Controversy in spin fluctuation mediated SC & even in the nesting properties)
Iron pnictides have attracted much attention due to its high T$_c$ superconductivity (SC).\cite{Kam2008}
A prime candidate for the pairing glue is the fluctuating antiferromagnetic (AF) order.
It has been argued that itinerant electrons form a spin density wave (SDW) via Fermi surface (FS) nesting,
and that fluctuating SDW causes the SC transition in the vicinity of the SDW phase boundary.\cite{Mazin2008}
In this scenario, a key ingredient to SC is the FS nesting property.
However, the FS nesting property of iron pnictides has been controversial.
Some systems are believed to possess good nesting properties,\cite{Tay2011,Qure2011,Put2011}
while some are not.\cite{Bor2010,Bryd2011}
Also there have been many theories that emphasize the role of local Fe $3d$ electrons.\cite{loc1,loc2,loc3}

%(To resolve the issue, exact determination of nesting properties is important:
Resolving such controversy requires accurate determination of FS topology and orbital characters.
Theoretical simulations based on a first principles method can provide such information.
However, the conventional density functional theory (DFT) method often fails to describe the electronic structure due to the significant
electron correlation effect in Fe-based superconductors.
Calculated bands have to be renormalized by approximately 2$\sim$4 to fit the experimentally measured band width,
and the predicted spin magnetic moment is about twice larger than the experimental value.
On the other hand, the dynamical mean field theory (DMFT) on top of the DFT showed consistent results  with
the measured bands, the anisotropy and the small magnetic moment.\cite{Craco2008,Yin2011,Lee2010,Hans2010,Yin2011a,HSJ}

%(Comparison with exp is essential and that is why Li111 is an important system)
Even though DMFT has been shown to work well, the comparison of calculated and measured band structures, especially the FS,
is still important to check the validity of the calculation. For an accurate comparison of theory and experiment,
LiFeAs is the most suitable system at present. The most studied 122 (isostructural to BaFe$_2$As$_2$) systems do not have neutral cleavage planes,
which affects surface sensitive techniques such as the angle resolved photoemission spectroscopy (ARPES).

%As a result, second derivatives are taken for ARPES data from 122 systems to determine band dispersions.
On the other hand, neutral cleavage surfaces of LiFeAs allow us
an accurate experimental determination of the band structure, FS topology and quasi-particle dynamics.\cite{Bor2010}
In this respect, DMFT correction on the band structure of LiFeAs in comparison with experimental data should provide a unique opportunity
to obtain accurate electronic structure information and resolve some of the issues on the origin of SC.

%(Purpose of the study)
For this reason, we performed both DMFT and ARPES studies on an intrinsic iron pnictide superconductor LiFeAs.
Our goal is to validate the accuracy of our DMFT method by comparing with ARPES results, and
to unravel the SC mechanism by analyzing the DMFT spectral function.
It will be shown that the FS nesting is marginally suppressed in LiFeAs due to the selective correlation effect of each Fe 3d orbital,
and that this gives rise to the electron pairing mediated by the spin-fluctuation (SF).

%(Description of calculation and experimental methods)
Simulation on LiFeAs is based on the fully self-consistent DFT
in combination with DMFT (DFT+DMFT) as implemented in WIEN2k.\cite{Blaha2001,Perdew1996,Kot2006,Haule2010}
The local self-energy due to the correlated Fe 3$d$ orbital is obtained with the
continuous time quantum Monte Carlo (CTQMC) impurity solver, where U=5.0 eV and J=0.8 eV are used.\cite{GW}
ARPES experiments were performed at HiSoR BL-9 and ALS BL-7 with similar conditions in Ref.\cite{arpes}.
Single crystals used in the experiments were synthesized by Sn-flux method.\cite{synth}

\begin{figure}[b]
\begin{center}
\includegraphics[width=8.5cm,height=10.0cm]{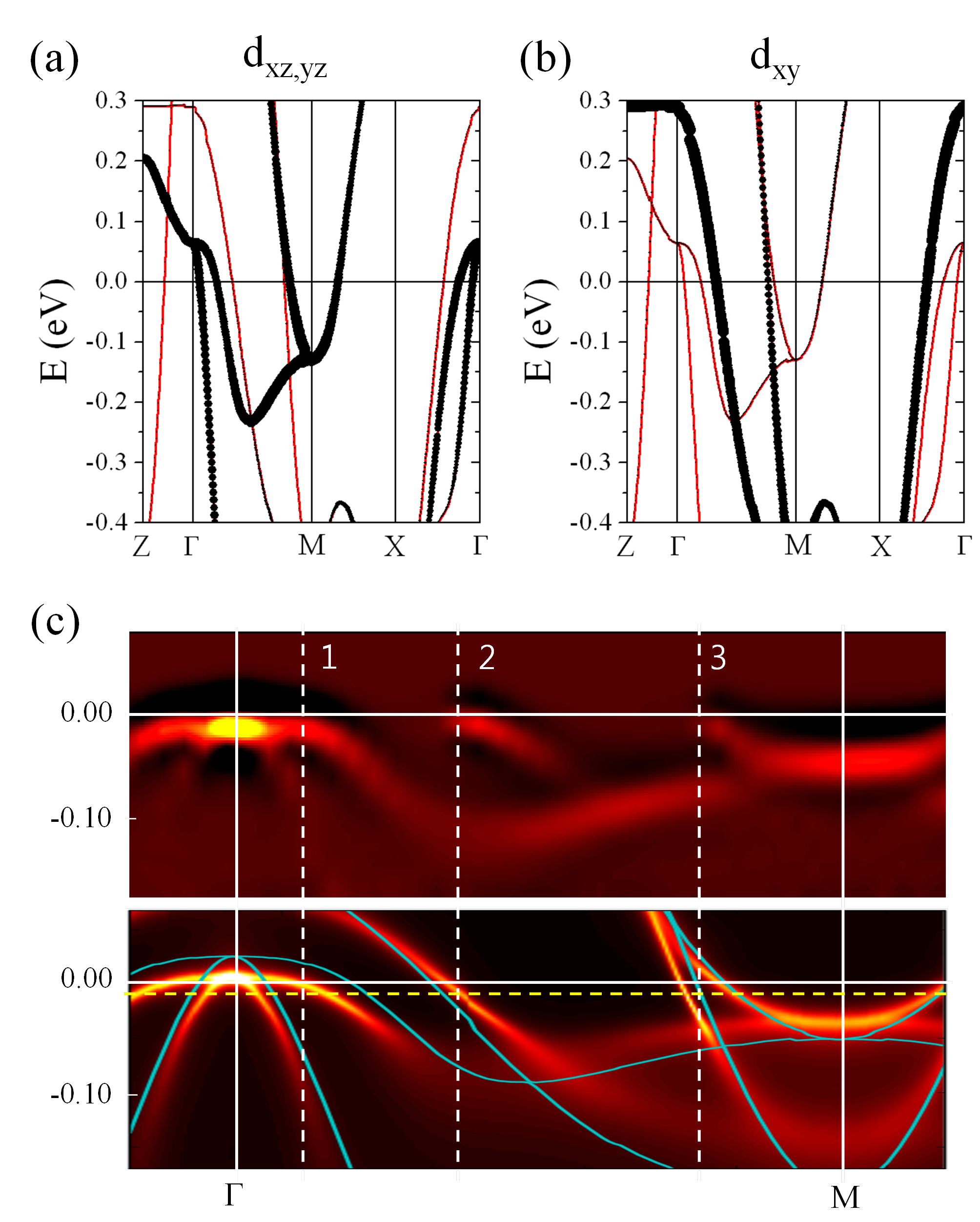}
\caption{\label{fig1} (color online)
DFT band structures (red lines) with (a) $d_{xz,yz}$ and (b) $d_{xy}$ contributions indicated
by the size of black circles.
(c) (upper panel) Second derivatives of ARPES data along the $\Gamma$-M direcition.
(lower panel) DFT+DMFT spectral functions in red color scale map, and DFT
band structures in green solid lines. DFT results are rescaled by an average renormlization
factor of 2.55.}
\end{center}
\end{figure}

% DMFT band and selective FS shifts
In Figs. 1(a) and 1(b), the DFT band structures are shown by thin gray lines. Overlayed on top of the bands are
the $d_{xz,yz}$ (Fig. 1(a)) and $d_{xy}$ (Fig. 1(b)) orbital contributions, respectively, as indicated
by the size of black circles.
Two small hole bands near the $\Gamma$ and one small electron band near $M$ are
mostly from $d_{xz,yz}$ orbitals, while the large hole and electron bands
are from $d_{xy}$ orbital.
Using the self-energy $\Sigma(\omega)$ obtained from the DFT+DMFT calculation,
we compute the momentum-resolved spectral function to inspect relative changes of those bands.
The result is shown in the lower panel of Fig. 1(c).
The DFT band structure in Fig. 1(c) is scaled by an average renormalization factor of 2.55 for comparison.
The most noticeable aspect of the data near $\Gamma$ is that two $d_{xz,yz}$ related inner hole bands are
{\it selectively} lowered, while a $d_{xy}$ related outer hole band is slightly shifted up, which will be discussed.

% comparison to ARPES
The calculated spectrum can be compared with the ARPES data in the upper panel of Fig. 1(c) presented in the same scale.
Three hole bands near $\Gamma$ are shown clearly in both theoretical and
experimental spectra.
Near the $M$ point, there is only one electron band in the ARPES, though
the DFT+DMFT predicts two bands crossing E$_F$.
The discrepancy can be ascribed to the matrix element effect.
Indeed, one can observe in the experimental FSs in Figs. 2(a) and 2(b) that two bands cross the E$_F$ near the $M$ point
where the orbital character varies with the Fermi wavevector.
The vertical dashed lines in Fig. 1(c) denote positions of experimental Fermi wavevectors.
Our theoretical predictions exhibit small deviation from experimental ones, especially for the outer electron pocket near M.
This might be due to the usual Li deficiency in the samples.
A better agreement is obtained if we lower the Fermi level by 0.01 eV
as indicated by the horizontal dashed line in Fig. 1(c) (4\% Li deficiency).
\cite{Pit2008} On the contrary, shift of the Fermi level alone does not give a good agreement between DFT (green sold lines) and experimental results.
Especially, the hole bands deviate even further.

\begin{figure}[b]
\begin{center}
\includegraphics[width=8.0cm]{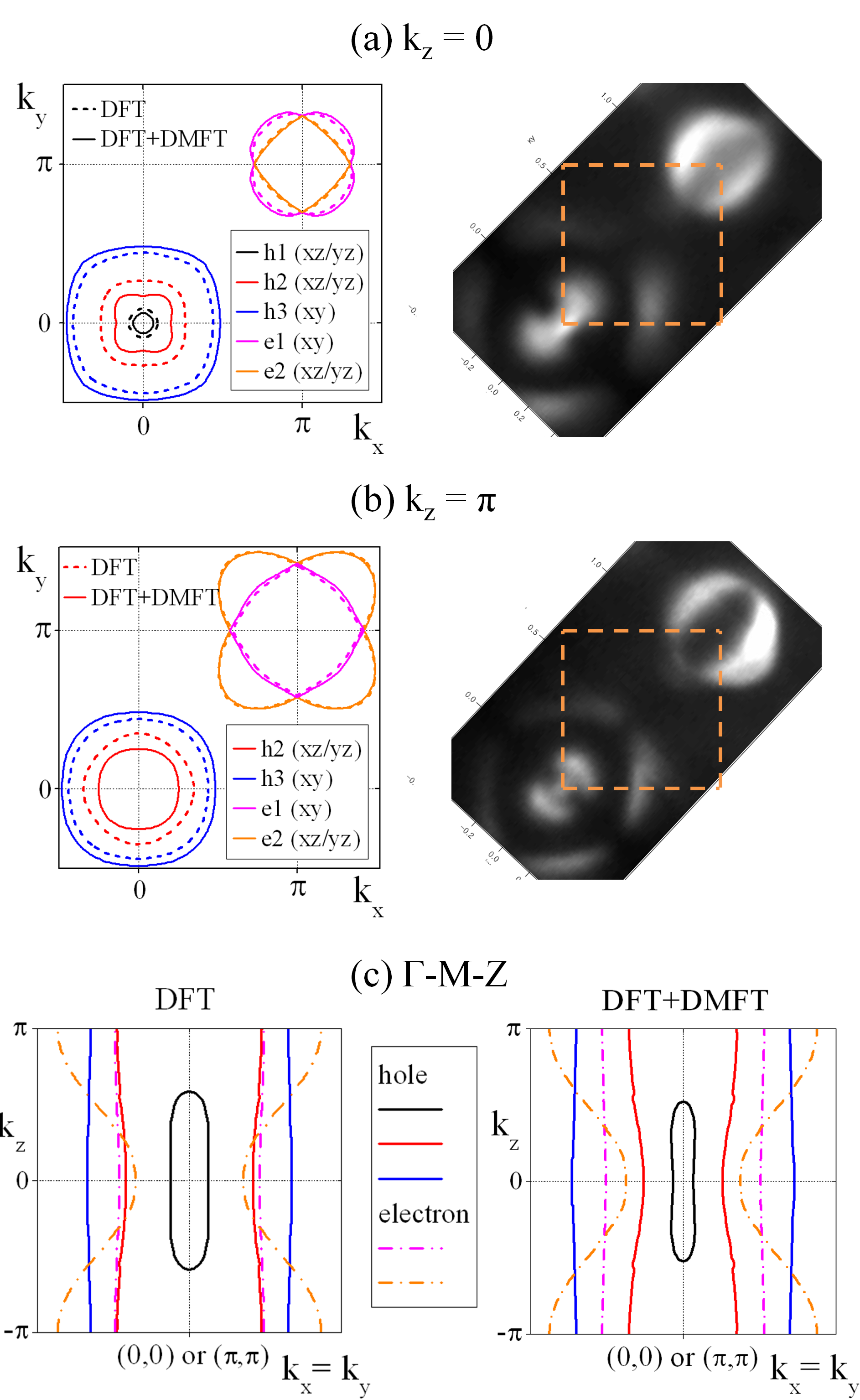}
\caption{\label{fig2} (color online)
DMFT (solid lines) and DFT (dashed) Fermi surfaces at (a) $k_z$=0, (b) $k_z=\pi$ are
compared with the experimental FSs. There are three hole pockets (h1, h2, h3) near $\Gamma$ and
two electron pockets (e1, e2) near $M$. The smallest hole pocket (h1) has strong $k_z$ dependence and does not appeat at $k_z=\pi$.
(c) $k_z$ dependence of FSs along the $\Gamma$-$M$-$Z$ cut. To check the nesting condition, electron pockets denoted by dot-dashed lines are shifted by ($\pi$,$\pi$) and plotted at $\Gamma$ on top of hole pockets (solid lines). DFT and DFT+DMFT results are shown in left and right panels, respectively.}
\end{center}
\end{figure}

%FS result (figure 2)
The change of the FS size due to the correlation effect affects the nesting condition. Shown in Figs. 2(a) and 2(b) are FSs at $k_z$=0 and $\pi$, respectively.
One can see that $xz/yz$-driven hole FSs shrink while the $xy$-driven hole FS
expands compared to the DFT results. Similar results have been recently reported for various Fe-based superconductors.\cite{Yin2011,Yin2011a}
Our DFT+DMFT results also agree well with the experimental results on the right.
Meanwhile, we do not see a sizable difference in the electron FS size for the two methods, which results in
supression of FS nesting between hole and electron pockets in the DFT+DMFT results.

To check the nesting conditions, we show in Fig. 2(c) overlayed hole and electron FSs in the $\Gamma$-M-Z plane.
While DFT result exhibits a good nesting between the second hole pocket
and an electron pocket (left panel), nesting becomes poor for the DFT+DMFT result (right panel).
Such difference in the nesting property can influence the quantum transition behavior dramatically,
as will be shown later.

\begin{figure}[b]
\begin{center}
\includegraphics[width=6.0cm]{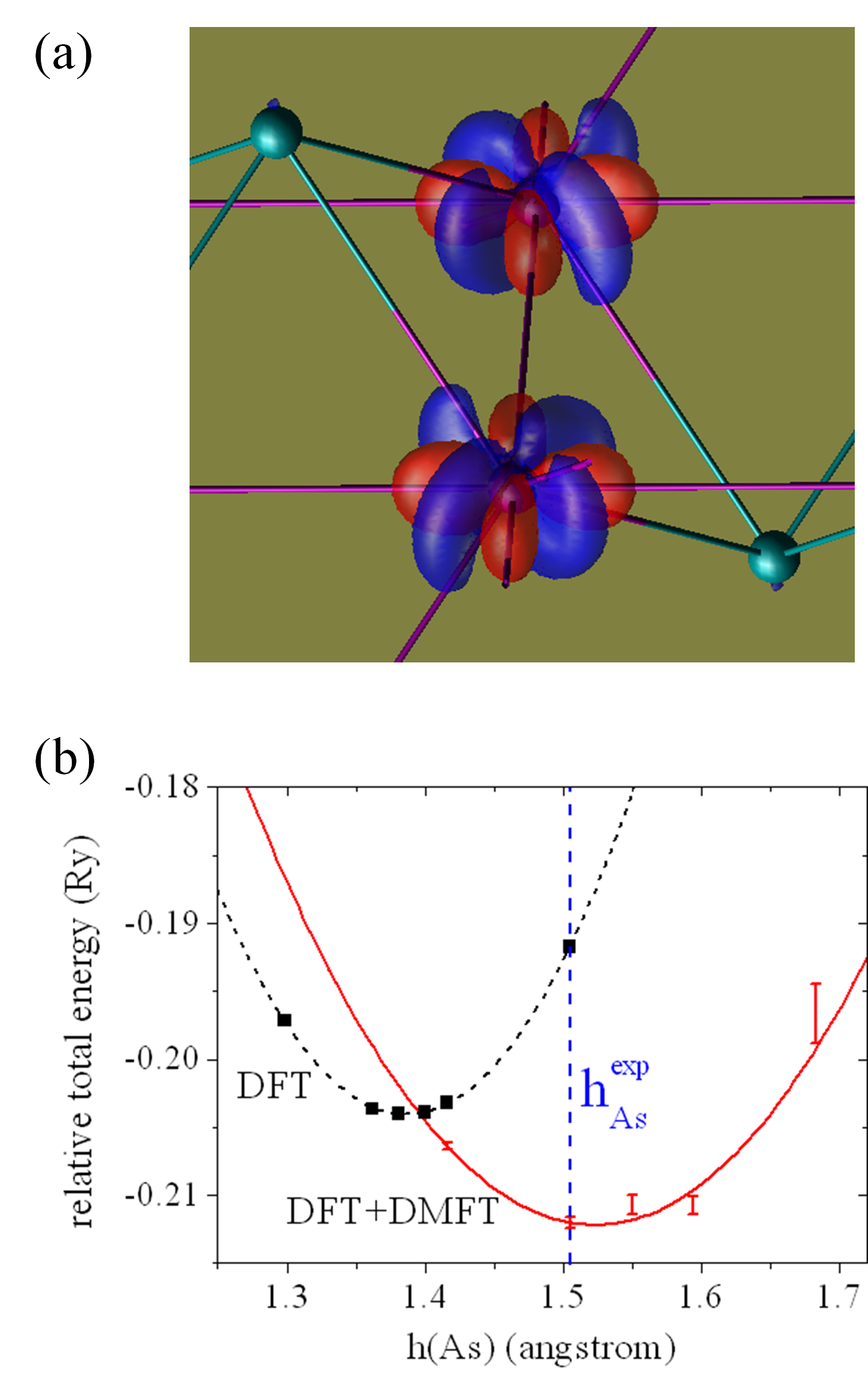}
\caption{\label{fig3} (color online)
(a) Isosurface plot of electron density difference between DFT and DFT+DMFT results.
The red (bule) color means increase (decrease) of 0.003 e/{\AA}$^3$ upon the DMFT calculation.
The cartesian $x$ or $y$ axis is chosen along the Fe-Fe bonds, so the increase is
related mainly to $d_{xz}$ or $d_{yz}$ orbital of Fe, while the decrease to $d_{xy}$.
(b) Calculated DFT and DFT+DMFT total energies against As height relative to the Fe plane.}
\end{center}
\end{figure}

% electron density
A natural question is on the mechanism of selective FS shift that increases (decreases) the $xz/yz$ ($xy$) hole FS size and affects
the nesting condition. We show that it is related to
the charge transfer among Fe-3$d$ orbitals which is done during the fully self-consistent DFT+DMFT steps.
We have checked the spectral function calculated by the DMFT self energy without
charge self consistency, and could not observe clear energy shifts as in the fully self-consistent case.
Fig. 3(a) shows the variation of the DMFT charge density from the DFT result.
The electron density increases along Fe-Fe bonds (red color) whereas it decreases along Fe-As bonds (blue).
Since the cartesian $x$- and $y$-axes are chosen along Fe-Fe bonds in our calculation,
the $d_{xz/yz}$ orbital occupancy is increased while the $d_{xy}$ orbital occupancy is decreased.
This is consistent with the downshift of $d_{xz/yz}$-related bands and upshift of $d_{xy}$-related bands shown
in Fig. 1.

Charge transfer from $xy$ to $xz/yz$ orbital can be understood to be caused by the difference in hybridization
magnitude, $t$ of each orbital. Under the common Coulomb interaction $U$,
the renormalization factor, $1/Z \propto U/t$, varies from 2.1 for $z^2$ and $x^2-y^2$ to 3.9 for $xy$. $xz/yz$ has an intermediate value of 2.9.
Therefore, the $xy$ orbital experiences a smaller hybridization (or larger localization) compared to $xz/yz$.
It means a larger Coulomb energy cost, which favors less occupation in $xy$.
Since $t_{2g}$ orbitals mainly contribute to near $E_F$ states, $xy$ electrons will be transfered to $xz/yz$ orbitals.

% energy calculation details
The charge redistribution due to the electron correlation also substantially contributes to bonding properties.
The decrease in the electron density along the Fe-As bond direction makes the Fe-As bond weaker,
which induces a longer Fe-As bond or higher arsenic height compared to the DFT estimation.
In Fig. 3(b), we show the calculated total energy as a function of As height, h(As), by DFT and DFT+DMFT methods.
Compared to the experimental value of h(As)$_{exp}$=1.5  {\AA},\cite{Tapp2008}
the DFT method predicts a position 0.1 {\AA} lower, which indicates overestimation of the binding in the DFT method.
Meanwhile, the DFT+DMFT estimation is in good agreement with h(As)$_{exp}$.

\begin{figure}[t]
\begin{center}
\includegraphics[width=8.5cm]{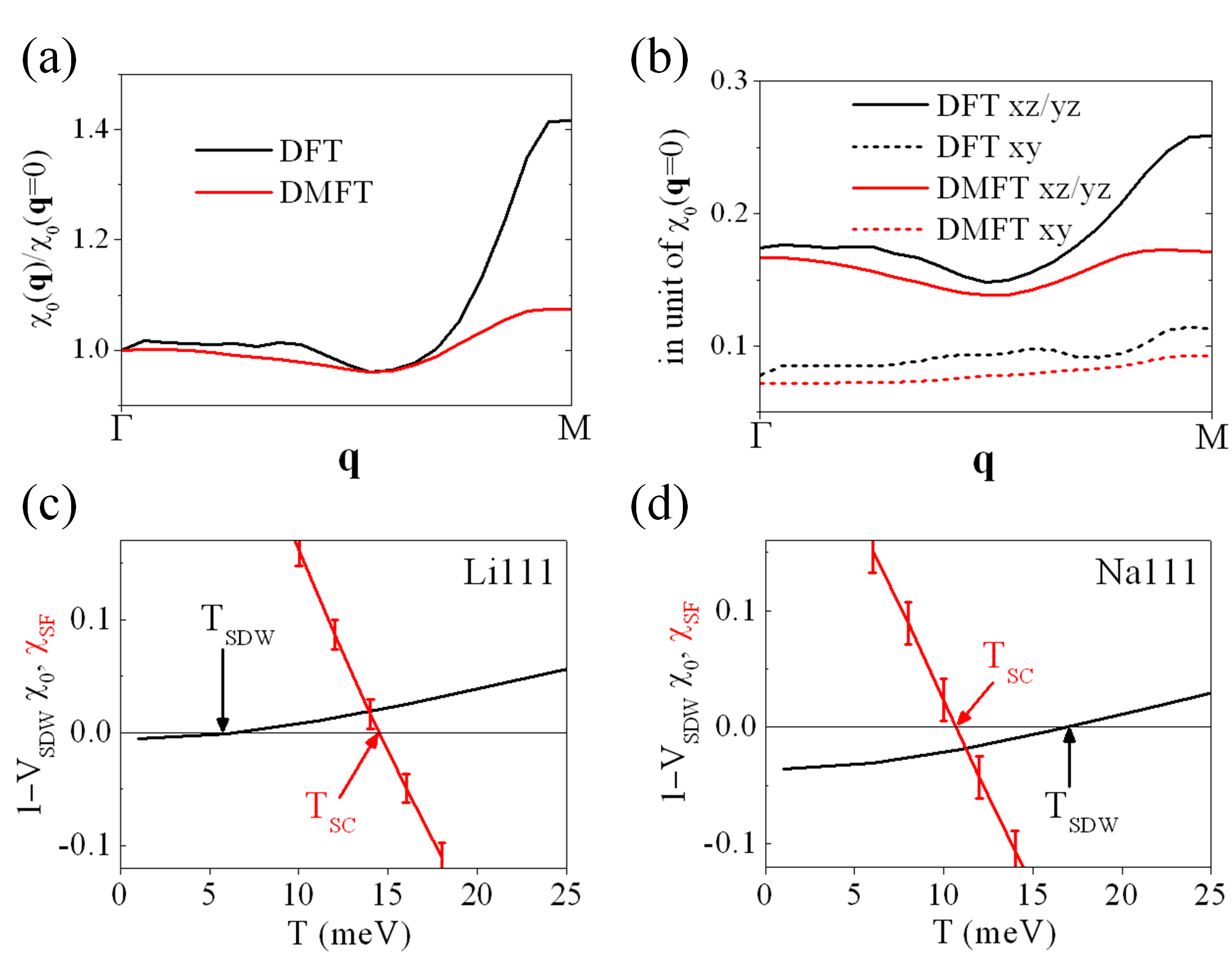}
\caption{\label{fig4} (color online)
(a) Bare susceptibilities $\chi_0(\mathbf{q})$ calculated by using the DMFT and DFT methods, normalized by
$\chi_0(\mathbf{q}=0)$.
(b) Intra-orbital scattering contribution to $\chi_0(\mathbf{q})$.
(c) and (d) SDW and SC instability plots versus temperature for LiFeAs (Li111) and NaFeAs (Na111), respectively.
}
\end{center}
\end{figure}

%FS nesting, chi0(q)
As shown above, our DFT+DMFT results are in good agreement with the experimental measurements and
shows that the nesting is marginally established. This suggests that LiFeAs is
located near the SDW boundary and that the SF can be the pairing glue for the SC.
In order to check such possibility, we have performed quantitative analysis of the nesting-driven instabilities
by utilizing the calculated spectral function $A(\mathbf{k},\epsilon)$.
In Fig. 4(a), we compare the bare susceptibilities $\chi_0(\mathbf{q})$.
The nesting enhancement at $\mathbf{q}={\rm M}$ is much suppressed
by including the correlation effect. It is mainly due to the size mismatch between hole and electron FSs as shown in Fig. 2(c).
Since the $A(\mathbf{k},\epsilon)$ is mostly from $xz/yz$ and $xy$ orbitals,
we focus on their contributions to $\chi_0(\mathbf{q})$. In addition, only the
intra-orbital scattering is considered because it gives the dominant contribution to the SC.\cite{Kuroki}
In Fig. 4 (b), the $xz/yz$ intra-orbital contribution is significantly larger (about twice)
than that of $xy$. It is because there are more $xz/yz$ orbital derived states
near the Fermi level. The shape of $\chi_0(\mathbf{q})$ in Fig. 4(a)
is also well reproduced by the scattering between $xz/yz$ dominant states.
Therefore, the FS nesting can be understood to be effectively between two $xz,yz$ derived bands (hole and electron bands near $\Gamma$ and M, respectively).
Note that the smallest hole pocket at $\Gamma$ has a negligible effect on the nesting property.

Suppressed nesting predicted by the DMFT spectrum implies that the SDW phase is also suppressed and
the SF driven SC ordering becomes more probable.
To have a SDW phase, the following condition is satisfied for a given nesting vector $\mathbf{Q}$.
\begin{eqnarray}
1<V_{SDW}\chi_0(\mathbf{Q},T)
\end{eqnarray}
A similar condition follows for the SF-driven SC:
\begin{eqnarray}
1< \frac{1}{\beta \pi^4} \sum_{\mathbf{q}} \left [ V_{SF}(\mathbf{q}) \right ]^2 \chi_{SF}(\mathbf{q}),
\end{eqnarray}
where $V_{SF}(\mathbf{q})=V_{SDW}^2\chi_0(\mathbf{q})/\left [ 1-V_{SDW}\chi_0(\mathbf{q})\right ]$ denotes
the singlet interaction channel, and $\chi_{SF}(\mathbf{q})$ is the associated SF susceptibility.
Using the calculated DMFT $xz/yz$ spectra, we compute $\left [1-V_{SDW}\chi_0 \right ]$ and $\chi_{SF}$ at the nesting
vector $\mathbf{Q}=M$ as a function of temperature in order to check the leading order between SC and SDW.
Fig. 4(c) shows the critical T$_{SDW}$ with the condition in Eq. (1).
Also T$_{SC}$ is estimated from the function of $\chi_{SF}(\mathbf{Q})$ using the condition of Eq. (2).
Note that $\chi_{SF}(\mathbf{q})$ is a smooth monotonic function and does not need to be positive as shown in Fig. 4(c).
Since $V_{SF}(\mathbf{q})$ shows a singular behavior near the SDW ordering for given $\mathbf{Q}$ and T$_{SDW}$, the condition in Eq. (2) can be always realized when $\chi_{SF}(\mathbf{q})$ is positive.
So T$_{SC}$ can be defined where $V_{SF}(\mathbf{Q})$ becomes positive as decreasing temperature.
%Indeed $\chi_{SF}(\mathbf{Q})$ is negative at high temperature but monotonically increases to become positive where T$_{SC}$ is defined.
In LiFeAs, T$_{SC}$ is always estimated to be higher than T$_{SDW}$ as shown in Fig. 4 (c).

For comparison, we show in Fig. 4(d)  the SC and SDW instabilities of NaFeAs which has a SDW ground state.
One can see that T$_{SDW}$ is higher than T$_{SC}$ in NaFeAs, in agreement with experiments.
Such difference between NaFeAs and LiFeAs comes from the fact that the FS nesting is stronger in NaFeAs (in the DMFT calculation), which enhances the SDW instability and increases the T$_{SDW}$.
On the other hand, SF is suppressed due to the stable SDW phase and thus T$_{SC}$ is decreased.
As a result, our DFT+DMFT method correctly predicts both SC and SDW ground states of LiFeAs and NaFeAs, respectively.
Note that the GGA spectrum does not give a SC transition because both compounds show good nesting features, resulting in a stable SDW ground state.

In summary we have shown that the inclusion of electron correlation effect is essential to correctly describe the orbital occupation, FS sizes, and As height, all which are important factors for unraveling the SC pairing mechanism.\cite{Hashimoto2012}
Our fully self-consistent DFT+DMFT calculation has accurately described the marginal FS nesting of LiFeAs observed in ARPES experiment. Using the DFT+DMFT spectrum, we also have successfully reproduced the SF-mediated SC in LiFeAs and SDW ground state in NaFeAs.

\begin{acknowledgments}
We acknowledge useful discussions with Y. Matsuda and K. Kuroki.
This work was supported by the NRF (No. 2010-0006484, 2010-0026762, 2011-0010186, 2011-0074542),
WCU through NRF (No. R32-2008-000-10180-0), and KISTI (KSC-2011-G3-02).
ARPES work was supported by NRF (No. 2010-0018092).
The work at SNU was financially supported by the National Creative Research Initiative (2010-0018300).
\end{acknowledgments}


\begin{thebibliography}{00}
\bibitem{Kam2008} Y. Kamihara \emph{et al.}, J. Am. Chem. Soc. {\bf 130}, 3296 (2008).
\bibitem{Mazin2008} I. Mazin \emph{et al.}, Phys. Rev. Lett. {\bf 101}, 057003 (2008).
\bibitem{Qure2011} N. Qureshi \emph{et al.}, Phys. Rev. Lett. {\bf 108}, 117001 (2012).
\bibitem{Tay2011} A. E. Taylor \emph{et al.}, Phys. Rev. B {\bf 83}, 220514 (2011).
\bibitem{Put2011} C. Putzke \emph{et al.}, Phys. Rev. Lett. {\bf 108}, 047002 (2012).
\bibitem{Bor2010} S. V. Borisenko \emph{et al.}, Phys. Rev. Lett. {\bf 105}, 067002 (2010).
\bibitem{Bryd2011} P. M. R. Brydon \emph{et al.}, Phys. Rev. B {\bf 83}, 060501 (2011).
\bibitem{loc1} P. Goswami \emph{et al.}, Phys. Rev. B {\bf 84}, 155108 (2011).
\bibitem{loc2} R. Applegate \emph{et al.}, Phys. Rev. B {\bf 81}, 024505 (2010).
\bibitem{loc3} A.  L. Wysocki \emph{et al.}, Nature Phys. {\bf 7}, 485 (2011).
\bibitem{Craco2008} L. Craco \emph{et al.}, Phys. Rev. B {\bf 78}, 134511 (2008).
\bibitem{Yin2011} Z. P. Yin \emph{et al.}, Nature Phys. {\bf 7}, 294 (2011).
\bibitem{Lee2010} H. Lee \emph{et al.}, Phys. Rev. B {\bf 81}, 220506 (2010).
\bibitem{Hans2010} P. Hansmann \emph{et al.}, Phys. Rev. Lett. {\bf 104}, 197002 (2010).
\bibitem{HSJ} H. S. Ji, G. Lee, J. H. Shim, Phys. Rev. B {\bf 84}, 054542 (2011).
\bibitem{Yin2011a} Z. P. Yin \emph{et al.}, Nature Mater. {\bf 10}, 932 (2011).
\bibitem{Blaha2001} P. Blaha, K. Schwarz, G. K. H. Madsen, D. Kvasnicka, and J. Luitz, WIEN2k, ISBN 3-9501031-1-2 (2001).
\bibitem{Perdew1996} J. P. Perdew, K. Burke, and M. Ernzerhof, Phys. Rev. Lett. {\bf 77}, 3865 (1996).
\bibitem{Kot2006} G. Kotliar, S. Y. Savrasov, K. Haule, V. S. Oudovenko, O. Parcollet, and C. A. Marianetti, Rev. Mod. Phys. {\bf 78}, 865 (2006).
\bibitem{Haule2010} K. Haule, C.-H. Yee, K. Kim, Phys. Rev. B. {\bf 81}, 195107 (2010).
\bibitem{GW} A. Kutepov \emph{et al.}, Phys. Rev. B {\bf 82}, 045105 (2010).
\bibitem{arpes} Y. Kim \emph{et al.}, Phys. Rev. B {\bf 83}, 064509 (2011).
\bibitem{synth} B. Lee \emph{et al.}, Europhys. Lett. {\bf 91}, 67002 (2010).
\bibitem{Pit2008} M. J. Pitcher  \emph{et al.}, Chem. Commun.  {\bf }, 5918 (2008).
\bibitem{Tapp2008} J. H. Tapp \emph{et al.}, Phys. Rev. B {\bf 78}, 060505(R) (2008).
\bibitem{Kuroki} K. Kuroki \emph{et al.}, Phys. Rev. B {\bf 79}, 224511 (2009).
\bibitem{Hashimoto2012} K. Hashimoto \emph{et al.}, Phys. Rev. Lett. {\bf 108}, 047003 (2012).
\end{thebibliography}
\end{document}

\title{Supplementary information: Orbital Selective Fermi surface shifts and mechanism of high T$_c$ superconductivity in correlated {\it A}FeAs ({\it A}=Li,Na)}

\author{Geunsik Lee }
\email{maxgeun@postech.ac.kr}
\affiliation{Department of Chemistry, Pohang University of Science and Technology, Pohang 790-784, Republic of Korea}
\affiliation{Center for Superfunctional Materials, Department of Chemistry, Pohang University of Science and Technology, Pohang 790-784, Republic of Korea}
\author{Hyo Seok Ji}
\affiliation{Department of Chemistry, Pohang University of Science and Technology, Pohang 790-784, Republic of Korea}
\author{Yeongkwan Kim}
\affiliation{Institute of Physics and Applied Physics, Yonsei University, Seoul, Republic of Korea }
\author{Changyoung Kim}
\affiliation{Institute of Physics and Applied Physics, Yonsei University, Seoul, Republic of Korea }
\author{Kristjan Haule}
\affiliation{Department of Physics and Astronomy, Rutgers University, Piscataway, New Jersey 08854, United States}
\author{Gabriel Kotliar}
\affiliation{Department of Physics and Astronomy, Rutgers University, Piscataway, New Jersey 08854, United States}
\author{Bumsung Lee}
\affiliation{Department of Physics and Astronomy, Seoul National University, Seoul 151-747, Republic of Korea}
\author{Seunghyun Khim}
\affiliation{Department of Physics and Astronomy, Seoul National University, Seoul 151-747, Republic of Korea}
\author{Kee Hoon Kim}
\affiliation{Department of Physics and Astronomy, Seoul National University, Seoul 151-747, Republic of Korea}
\author{Kwang S. Kim}
\affiliation{Center for Superfunctional Materials, Department of Chemistry, Pohang University of Science and Technology, Pohang 790-784, Republic of Korea}
\affiliation{Department of Physics, Pohang University of Science and Technology, Pohang 790-784, Republic of Korea}
\author{Ki-Seok Kim}
\affiliation{Asia Pacific Center for Theoretical Physics, Pohang University of Science and Technology, Pohang 790-784, Republic of Korea}
\affiliation{Department of Physics, Pohang University of Science and Technology, Pohang 790-784, Republic of Korea}
\author{Ji Hoon Shim}
\email{jhshim@postech.ac.kr}
\affiliation{Department of Chemistry, Pohang University of Science and Technology, Pohang 790-784, Republic of Korea}
\affiliation{Department of Physics, Pohang University of Science and Technology, Pohang 790-784, Republic of Korea}

\date{\today}

\maketitle

\section{An effective two-band model Hamiltonian}

One of two bands is a hole band centred at (0,0) and the other is an electron band at ($\pi$,$\pi$).
Those are driven by Fe $d_{xz/yz}$ orbital, which gives rise to intra-orbital scattering
The effective Hamiltonian is given below.
\begin{eqnarray}
&& H=H_0 + H_{\Delta} + H_{M} \\
&& H_0 = \sum_{\mathbf{k},\alpha} \xi_c(\mathbf{k}) c_{\mathbf{k}\alpha}^\dagger c_{\mathbf{k}\alpha}
       + \sum_{\mathbf{p},\alpha} \xi_f(\mathbf{p}) f_{\mathbf{p}\alpha}^\dagger f_{\mathbf{p}\alpha} \\
&& \xi_c(\mathbf{k}) = E_c - \frac{k^2}{2 m_c} \\
&& \xi_f(\mathbf{p}) = E_f + \frac{p^2}{2 m_f} \\
\end{eqnarray}
$\xi_c(\mathbf{k})$ and $\xi_f(\mathbf{p})$ represent dispersions of hole band $c$ and electron band $f$,
where $\mathbf{k}$ is relative to (0,0) and $\mathbf{p}$ is relative to ($\pi$,$\pi$).

The SC pairing interaction $H_\Delta$ is characterized by the coupling constant $V^{SC}$.
\begin{eqnarray}
&& H_{\Delta} = \frac{1}{2} \sum_{\mathbf{k},\mathbf{p}} V_{\alpha\beta\beta'\alpha'}^{cf}(\mathbf{k},\mathbf{p}) \left ( c_{\mathbf{k}\alpha}^\dagger c_{-\mathbf{k}\beta}^\dagger f_{-\mathbf{p}\beta'} f_{\mathbf{p}\alpha'} + f_{\mathbf{k}\alpha}^\dagger f_{-\mathbf{k}\beta}^\dagger c_{-\mathbf{p}\beta'} c_{\mathbf{p}\alpha'} \right) \\
&& V_{\alpha\beta\beta'\alpha'}^{cf}(\mathbf{k},\mathbf{p}) = V_{\mathbf{k},\mathbf{p}}^{SC} (i\mathbf{\sigma}^y)_{\alpha\beta} (i\mathbf{\sigma}^y)_{\beta'\alpha'}^\dagger \\
\end{eqnarray}

The SDW interaction $H_M$ is characterized by the coupling constant $V^{SDW}$.
\begin{eqnarray}
&& H_M = -\frac{1}{4} \sum_{\mathbf{p}'-\mathbf{p}=\mathbf{k}'-\mathbf{k}} V_{\alpha\beta\beta'\alpha'}^{SDW}(\mathbf{p}'\mathbf{p};\mathbf{k}\mathbf{k}')\left( f_{\mathbf{p}'\alpha}^\dagger c_{\mathbf{p}\beta}c_{\mathbf{k}\beta'}^\dagger f_{\mathbf{k}'\alpha'} + f_{\mathbf{-p}'\alpha}^\dagger c_{\mathbf{-p}\beta}c_{\mathbf{-k}\beta'}^\dagger f_{\mathbf{-k}'\alpha'} \right ) \\
&& V_{\alpha\beta\beta'\alpha'}^{SDW}(\mathbf{p}'\mathbf{p};\mathbf{k}\mathbf{k}') = V_{\mathbf{p}'\mathbf{p};\mathbf{k}\mathbf{k}'}^{SDW} \mathbf{\sigma}_{\alpha\beta}\cdot\mathbf{\sigma}_{\beta'\alpha'}^\dagger
\end{eqnarray}

\section{SDW instability}

We ignore the full momenta dependence of $V^{SDW}$ in Eq. (S10). So $ V_{\mathbf{p}'\mathbf{p};\mathbf{k}\mathbf{k}'}^{SDW}$ is assumed as a constant $V_{SDW}$.
Electron-hole scattering SDW susceptibility within RPA can be written like below.
\begin{eqnarray}
&&\chi_{SDW}(\mathbf{q})=\frac{V_{SDW}}{1-V_{SDW}\chi_0(\mathbf{q})} \\
&&\chi_0(\mathbf{q})=\frac{1}{\pi^2} \sum_\mathbf{p} \int d\epsilon\int d\epsilon' A_f(\mathbf{p}+\mathbf{q},\epsilon) A_c(\mathbf{p},\epsilon') \frac{f(\epsilon)-f(\epsilon')}{\epsilon-\epsilon'}
\end{eqnarray}

The spectral functions $A_f(\mathbf{p},\epsilon)$, $A_c(\mathbf{p},\epsilon)$ are computed from our DFT or DFT+DMFT results.
From obtained Green's function $G_{ij}(\mathbf{p},\epsilon)$, where $i,j$ represents band index,
the whole hole spectrum is computed by $(-1/\pi)\sum_{i<i_c} {\rm Im} G_{ii}(\mathbf{p},\epsilon)$.
Likewise the whole electron spectrum is $(-1/\pi)\sum_{i\geq i_c} {\rm Im} G_{ii}(\mathbf{p},\epsilon)$.
Here $i_c$ is a chosen band number which separates hole and electron spectra.
Finally obtained hole and electron spectra are projected into $xz/yz$ orbital of Fe.
Using this procedure, we show computed $A_f(\mathbf{p},\epsilon)$ and $A_c(\mathbf{p},\epsilon)$ in Fig. S1 along $\Gamma$-M direction near the Fermi level.
One can see clearly the hole and electron bands near $\Gamma$ and M, respectively.
Also $A_f$ and $A_c$ behave like quasiparticle, so they are applicable to the effective two-band model.
We note that one inner hole band among two $xz/yz$-drived bands in Fig. S1a has minor effect in drawing our main conclusion.

At $T=T_{SDW}$ the following condition is satisfied for a given nesting vector $\mathbf{Q}$.
\begin{eqnarray}
1=V_{SDW}\chi_0(\mathbf{Q},T=T_{SDW})
\end{eqnarray}

\begin{figure}[b]
\begin{center}
\includegraphics[width=14.cm]{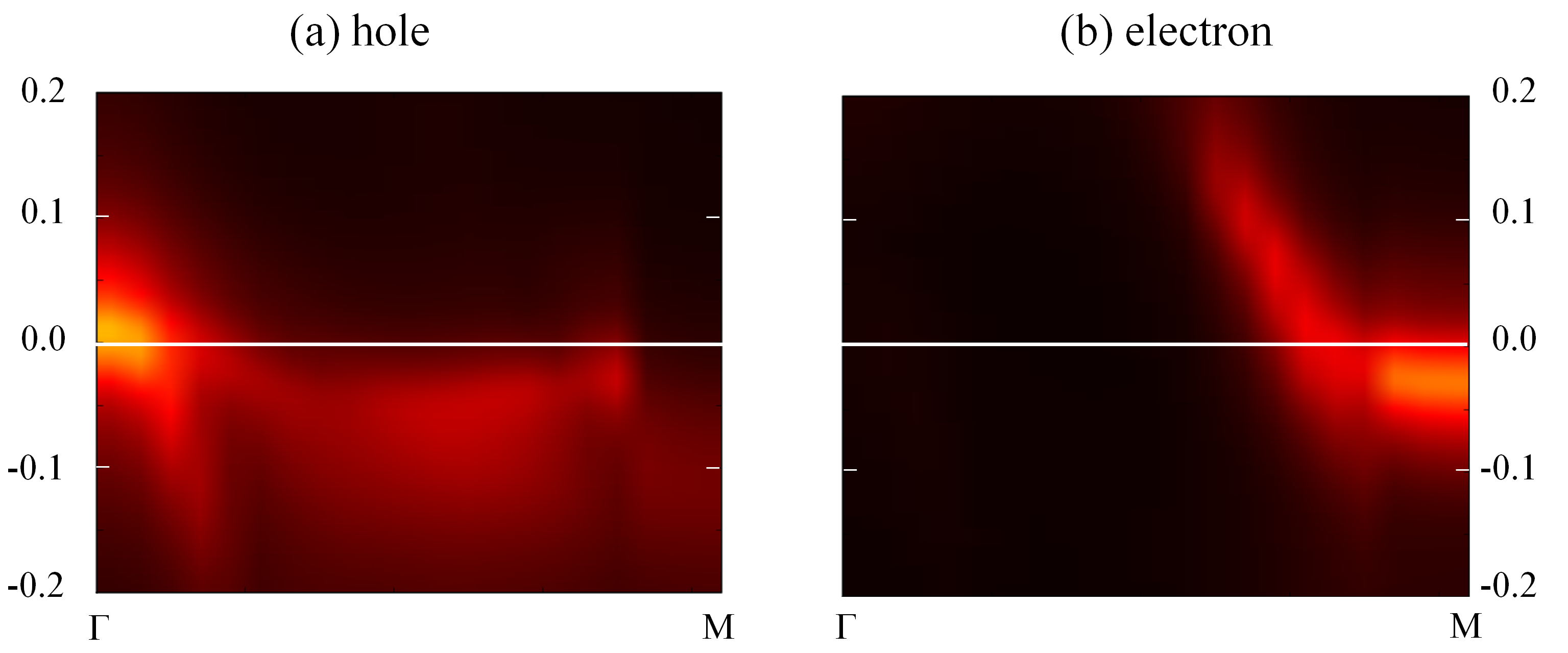}
\caption{ (color online)
(a) $A_c(\mathbf{k},\epsilon)$, (b) $A_f(\mathbf{k},\epsilon)$ of LiFeAs by DFT+DMFT method.
}
\end{center}
\end{figure}

\newpage

\section{Spin-fluctuation mediated superconductivity}

The singlet interaction channel is given by
\begin{eqnarray}
&& V_{SF}(\mathbf{q},i\Omega)=\chi(\mathbf{q},i\Omega)-V_{SDW}=\frac{V_{SDW}^2\chi_0(\mathbf{q},i\Omega)}{1-V_{SDW}\chi_0(\mathbf{q},i\Omega)}.
\end{eqnarray}

Assuming $s_{\pm}$ pairing symmetry, the SF-driven superconducting instability is derived using associated hole (electron) Green's function $G_{c(f)}$. It will be applicable to interacting spectral functions by DFT+DMFT.
\begin{eqnarray}
1=\frac{1}{\beta^2}\sum_{i\omega,i\Omega}\sum_{\mathbf{p},\mathbf{q}}
&& G_f(\mathbf{p},i\omega)   V_{SF}(\mathbf{q},i\Omega)G_c(\mathbf{p}+\mathbf{q},i\omega+i\Omega) \notag \\
&& \times G_f(-\mathbf{p},-i\omega) V_{SF}(\mathbf{q},i\Omega) G_c(-\mathbf{p}+\mathbf{q},-i\omega+i\Omega)
\end{eqnarray}
The right hand side can be transformed in a following form involving $A_{c(f)}$.
\begin{eqnarray}
&&\frac{1}{\pi^4}\frac{1}{\beta^2}\sum_{i\Omega}\sum_{\mathbf{p},\mathbf{q}}
\int d\epsilon \int d\epsilon' \int d\epsilon_1 \int d\epsilon_1'
\left [V_{SF}(\mathbf{q},i\Omega) \right ]^2
A_f(\mathbf{p},\epsilon) A_c(\mathbf{p}+\mathbf{q},\epsilon')
A_f(-\mathbf{p},\epsilon_1) A_c(-\mathbf{p}+\mathbf{q},\epsilon_1') \notag \\
&& \times \sum_{i\omega}
\frac{1}{\epsilon-i\omega} \frac{1}{\epsilon'-i\omega-i\Omega}
\frac{1}{\epsilon_1+i\omega} \frac{1}{\epsilon_1'+i\omega-i\Omega}
\end{eqnarray}
Also the last term is reexpressed by using Fermi-Dirac function $f(\epsilon)$.
\begin{eqnarray}
&& \frac{1}{\beta} \sum_{i\omega}
\frac{1}{\epsilon-i\omega} \frac{1}{\epsilon'-i\omega-i\Omega}
\frac{1}{\epsilon_1+i\omega} \frac{1}{\epsilon_1'+i\omega-i\Omega}
= \frac{1}{i\Omega+\epsilon-\epsilon'} \frac{1}{i\Omega+\epsilon_1-\epsilon_1'} \notag \\
&& \times \left [
\frac{f(\epsilon)+f(\epsilon_1)}{\epsilon+\epsilon_1} -
\frac{f(\epsilon')+f(\epsilon_1)}{-i\Omega+\epsilon'+\epsilon_1} -
\frac{f(\epsilon)+f(\epsilon_1')}{-i\Omega+\epsilon+\epsilon_1'} +
\frac{f(\epsilon')+f(\epsilon_1')}{-2i\Omega+\epsilon'+\epsilon_1'} \right ]
\end{eqnarray}
By using $V_{SF}(\mathbf{q},i\Omega) \approx V_{SF}(\mathbf{q},i\Omega=0)$, thus $V_{SF}(\mathbf{q},i\Omega) \rightarrow V_{SF}(\mathbf{q})$, we finally obtain a following form
\begin{eqnarray}
&& 1=\frac{1}{\beta\pi^4}\sum_{\mathbf{p}}\sum_{\mathbf{q}}
\int d\epsilon \int d\epsilon' \int d\epsilon_1 \int d\epsilon_1'
\left [ V_{SF}(\mathbf{q}) \right ]^2
A_f(\mathbf{p},\epsilon) A_c(\mathbf{p+q},\epsilon') A_f(\mathbf{-p},\epsilon_1) A_c(\mathbf{-p+q},\epsilon_1') \notag \\
&& \frac{1}{\epsilon-\epsilon'}\frac{1}{\epsilon_1-\epsilon_1'}
\left \{ \frac{f(\epsilon)+f(\epsilon_1)}{\epsilon+\epsilon_1}
-\frac{f(\epsilon')+f(\epsilon_1)}{\epsilon'+\epsilon_1}
-\frac{f(\epsilon)+f(\epsilon_1')}{\epsilon+\epsilon_1'}
+\frac{f(\epsilon')+f(\epsilon_1')}{\epsilon'+\epsilon_1'}
\right \}
\end{eqnarray}
We simplify the above expression as below.
\begin{eqnarray}
&&1 =\frac{1}{\beta\pi^4}\sum_{\mathbf{q}} \left [ V_{SF}(\mathbf{q}) \right ]^2
\chi_{SF}(\mathbf{q})
\end{eqnarray}

Computation of Eq. (S18) involves sixth-order integration including BZ summation, which is very challenging.
For the fourth order integration over the energy, we use normal Monte-Carlo sampling method.
Important parameters that we need to care for good accuracy are
(i) the number of $\mathbf{p}$ or $\mathbf{q}$ points,
(ii) the number of MC sampling points for a given energy integration range, i.e. energy window,
$\left | \epsilon -\mu \right | \le$ $E_w$.
The BZ integration over $\mathbf{p}$ or $\mathbf{q}$ is done with small number of points as possible.
So we choose the minimal grid size along z-direction that can give reasonable results, which is two.
Also it will be seen that the summation over $\mathbf{q}$ is not necessary to judge the SC transition above SDW temperature.
The $E_w$ is in principle to be infinite. But we choose such $E_w$ large enough to include most spectra of specific
hole and electron bands. So this parameter is dependent on a problem.

\section{The existence of T$_C$ above T$_{SDW}$}

Noting that $V_{SF}(\mathbf{q})$ is singular at the SDW nesting vector $\mathbf{Q}$,
we express Eq. S19 as below .
\begin{eqnarray}
{\beta\pi^4} = \frac{1}{N} \left [ V_{SC}(\mathbf{Q}) \right ]^2
\chi_{SC}(\mathbf{Q})+ \frac{N-1}{N} \sum_{\mathbf{q}\ne \mathbf{Q}} \left [ V_{SC}(\mathbf{q}) \right ]^2
\chi_{SC}(\mathbf{q})
\end{eqnarray}
Here $N$ is the number of $\mathbf{q}$ points in the full Brillouin zone, which is supposed to be finite number.

We suppose following two conditions,
(i) $\chi_{SC}(\mathbf{q}) > 0$ only at $\mathbf{Q}$,
(ii) $\frac{d \chi_{SC}(\mathbf{Q},T)}{dT}<0$.
The second condition is naturally satisfied in order that physically meaningfull SC transtion be observed at low
temperature.
Also in Eq. S18, the FD function $\frac{f(\epsilon)+f(\epsilon_1)}{\epsilon+\epsilon_1}$ satisfies the second condition.

In Eq. S20, the second term at the right hand side is a negative quantity.
At the T$_{SDW}$, if $\chi_{SC}(\mathbf{Q}) > 0$, singular $V_{SC}(\mathbf{Q})$ is large enough to satisfy
the following condition.
\begin{eqnarray}
\left [ V_{SC}(\mathbf{Q}) \right ]^2
\chi_{SC}(\mathbf{Q}) > N {\beta\pi^4} - (N-1) \sum_{\mathbf{q}\ne \mathbf{Q}} \left [ V_{SC}(\mathbf{q}) \right ]^2
\chi_{SC}(\mathbf{q})
\end{eqnarray}

Thus, from the condition (ii), there exists a SC transition which satisfies the Eq. S19 above T$_{SDW}$.
This is also true for the case that $\chi_{SC}(\mathbf{q})$ involves positive quantities partially at some $\mathbf{q}$ points.

%\begin{thebibliography}{00}
%\bibitem{Kam2008} Y. Kamihara, T. Watanabe, M. Hirano, and H. Hosono, J. Am. Chem. Soc. {\bf 130}, 3296 (2008).
%\end{thebibliography}